\documentclass{iopart}

\usepackage[english]{babel}
\usepackage{epsfig}
\usepackage{color}

\begin{document}

\title{Charge-memory effect in a polaron model: equation-of-motion method for Green functions}

\author{Pino~D'Amico$^{1}$\footnote{email:pino.damico@physik.uni-regensburg.de},
Dmitry\,A.~Ryndyk$^{1}$, Gianaurelio~Cuniberti$^{2}$, and Klaus~Richter$^{1}$}

\address{$^{1}$Institute for Theoretical Physics, University of Regensburg,
D-93040 Regensburg, Germany \\ $^{2}$ Institute for Material Science and Max
Bergmann Center of Biomaterials, Dresden University of Technology, D-01062 Dresden,
Germany}

\begin{abstract}
We analyze a single-level quantum system placed between metallic leads and strongly
coupled to a localized vibrational mode, which models a single-molecule junction or
an STM setup. We consider a polaron model describing the interaction between
electronic and vibronic degrees of freedom and develop and examine different
truncation schemes in the equation-of-motion method within the framework of
non-equilibrium Green functions. We show that upon applying gate or bias voltage, it
is possible to observe charge-bistability and hysteretic behavior which can be the
basis of a charge-memory element. We further perform a systematic analysis of the
bistability behaviour of the system for different internal parameters such as the
electron-vibron and the lead-molecule coupling strength.
\end{abstract}

PACS: 85.65.+h, 73.23.-b, 73.63.-b, 73.40.Gk

\maketitle

\markboth{Charge-memory effect in a polaron model: EOM method for Green functions
}{Charge-memory effect in a polaron model: EOM method for Green functions}

\section{Introduction}

Within the field of single-molecule electronics
\cite{Reed00sciam,Nitzan03science,Joachim05pnas,Cuniberti05book}, beside
experimental progress with regard to vibrational properties and their signatures in
transport \cite{Park00nature,Smit02nature,Yu04prl,Qiu04prl,Osorio07advmat}, related
phenomena such as switching, memory effects and hysteretic behavior in molecular
junctions have gained increasing importance and attention. Random and controlled
switching of single molecules
\cite{Loertscher07prl,Loertscher06Small,DelValle07naturenanotech}, as well as
conformational memory effects \cite{Choi06prl,Martin06prl,Liljeroth07science} have
been recently explored. Related to these effects, there is the so-called
charge-memory effect, that is basically a hysteretic behaviour of the
charge-voltage, respectively, current-voltage characteristics arising from the
interplay between the polaron shift and Franck-Condon blockade \cite{Koch05PRL} in
the presence of electron-vibron interaction. Several works have recently addressed
this interesting feature of molecular junctions, both experimentally
\cite{Repp04science,Olsson07prl} and theoretically
\cite{Alexandrov03prb,Galperin05nanolett,Mitra05prl,Mozyrsky06prb,Ryndyk08preprint}.

\begin{figure}[t]
\begin{center}
\vskip 0.5cm\epsfxsize=0.5\hsize \epsfbox{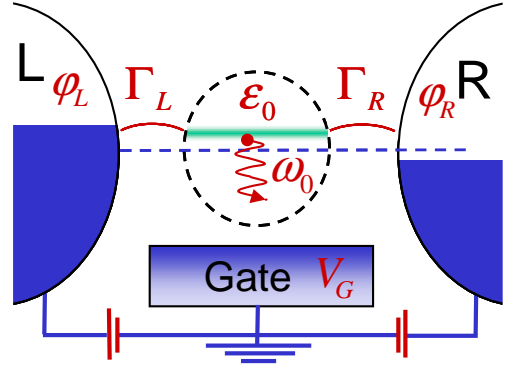}
\caption{(Color) Schematic representation of a single-level model system interacting with a vibronic mode and
coupled to left and right leads.}
\label{Paper-scheme}
\end{center}
\end{figure}

The charge-memory effect can be explained in the framework of a simple single-level
polaron model \cite{Galperin05nanolett,Ryndyk08preprint,Hewson74jjap,Hewson79jphysc}, where the
electronic state is coupled to a vibronic mode with frequency $\omega_0$, see the sketch
in Fig.~\ref{Paper-scheme}.  If the
energy of the {\em unoccupied} electron level {\em without} electron-vibron
interaction is $\epsilon_0$, the {\em occupied} (charged) state of the {\em
interacting system} will have the energy $\epsilon_1=\epsilon_0-\epsilon_p$, where
$\epsilon_p$ is so-called polaron shift (called also recombination energy). Neutral
and charged (polaron) states correspond to local minima of the potential energy
surface and get metastable, if the electron-vibron interaction is strong enough.
Upon applying an external voltage, one can change the state of this bistable system, an
effect that is accompanied by hysteretic charge-voltage and current-voltage curves.
In this model it is not necessary to include Coulomb interaction explicitly,
though one can straightforwardly incorporate charging effects.

It was alternatively suggested in Refs.~\cite{Mitra05prl,Mozyrsky06prb} that quantum
switching between bistable states rather results in telegraph noise at finite
voltage than in a memory effect. In a recent paper \cite{Ryndyk08preprint},
considering the problem in the weak molecule-to-lead coupling limit, {\em i.e.} for
the level width $\Gamma\ll\omega_0,\epsilon_p$, we considered the crossover between
these two pictures, if one takes into account the time scale of the switching
process. Indeed, the switching time $\tau$ between the two states of interest should
be compared with the characteristic time of the external voltage sweeping,
$\tau_s\sim V(t)/(dV(t)/dt)$. For $\tau\gg\tau_s$, quantum switching can be
neglected, and hysteresis can be observed, while in the opposite limit,
$\tau\ll\tau_s$, the averaging removes the hysteresis. In
Ref.~\cite{Ryndyk08preprint} we found that at large enough electron-vibron
interaction strenght $\lambda/\omega_0$ (where $\lambda$ is introduced below in
Eq.~(\ref{hamiltonian})), the switching time $\tau$ is exponentially large compared
to the inverse bare tunneling rate $1/ \Gamma$.

In this paper we consider the polaron problem in another limiting case,
$\omega_0\ll\Gamma<\epsilon_p$, corresponding to the regime where the
Born-Oppenheimer approximation holds true. In this situation of intermediate
molecule-to-lead coupling we address the case of time-independent applied voltages,
considering then the stationary problem and focusing on the properties of the two
states of interest. In this paper we consider the parameter ranges for which the
fluctuations between the two charge states of interest are negligible. It should be
noted that the mean-field solutions we consider are meta-stable, but they are
physical in the case of very long switching time $\tau$. Of course, the exact
equilibrium solution is a superposition of these two states. We assume that all
possible relaxation processes, with the exception of the switching between
metastable states, are much faster than $\tau$. Within the framework of
nonequilibrium Green functions
\cite{Meir92prl,Jauho94prb,Haug96book,Bruus04book,Ryndyk08review} we use an
equation-of-motion (EOM) approach which allows for studying the appearance of the
charge-memory effect addressed at different levels of approximation, starting with
the self-consistent Hartree level. The work is partly built on and further develops
ideas introduced in Refs.~\cite{Hewson74jjap,Hewson79jphysc,Galperin05nanolett} and
complements the results based on a master-equation approach in
Ref.~\cite{Ryndyk08preprint}.

The paper is organized as follows: In Sec.~2 we introduce the model and
the relevant quantities necessary to describe transport through the molecular
junction.  In Sec.~3 we outline the equation-of-motion method,
used for the calculation of nonequilibrium  Green functions, and discuss
different truncation schemes. In the final Sec.~4 we present and discuss
the results of our approach.

\section{Model and method}

\subsection{The single-level electron-vibron Hamiltonian}

We consider a three-terminal device with the central part given
by a single level interacting with a vibronic mode (see Fig.~\ref{Paper-scheme}).
Possible experimental implementations include metal-molecule-metal junctions or
STM spectroscopy of a single molecule on a conducting substrate. To describe such
a system we use the standard electron-vibron Hamiltonian
\begin{eqnarray} \label{hamiltonian}
\hat H & = & (\epsilon_{0}+e \varphi_{0})d^{\dagger}d+\omega_{0}a^{\dagger}a+\lambda
(a^{\dagger}+a)d^{\dagger}d + \nonumber \\
 & & +\sum_{i,k}\left[(\epsilon_{i,k}+e\varphi_{i})c^{\dagger}_{i,k}c_{i,k}
   +(V_{i,k}c^{\dagger}_{i,k}d+h.c.)\right] \, .
\end{eqnarray}
Here the terms in the first line describe the central system including one
electronic state with energy $\epsilon_{0}$, one vibronic state with frequency
$\omega_{0}$ and their mutual interaction with a coupling strength $\lambda$. The
second line in Eq.~(\ref{hamiltonian}) contains the Hamiltonian of the two leads
with independent-particle states and the tunneling between the leads and the central
region via the couplings $V_{i,k}$.  The index $i$ denotes the left and right leads,
while $k$ labels the electronic states of electrons in the leads.
Below it proves furthermore convenient to employ the
vibronic "position" and "momentum" operators
\begin{equation}
x=a^\dagger+a \quad ; \quad  p=a^\dagger-a \, .
\end{equation}

The energy level $\epsilon_0$ in Eq.~(\ref{hamiltonian}) can be shifted through the
gate voltage $V_G$. We choose as reference energy $\epsilon_{0} = 0$ for $V_G=0$ and
assume a linear capacitive coupling, $\epsilon_0= \alpha eV_G$ putting $\alpha=1$.
Note that the presence of a bias voltage $V_B=\varphi_{L}-\varphi_{R}$ can also
change the energy of the electronic level via the parameter
$\varphi_{0}=\varphi_{R}+\eta V_B$, where  $0 < \eta < 1$ describes the symmetry of
the voltage drop across the junction: $\eta=0$ corresponds to the completely
asymmetric case, while $\eta=0.5$ stands for the symmetric case. Thus in all
approximations the bias and gate voltages are taken into account twofold, through
the potentials of the leads, which can be chosen e.g. as $\varphi_L=V_B/2$,
$\varphi_R=-V_B/2$, and the effective energy of the level, which is correspondingly
$\tilde\epsilon_0=\epsilon_0+e\varphi_0=eV_G+e(\eta-0.5) V_B$. From this expression
for $\tilde{\epsilon}_0$ it follows that in the case of asymmetric bias-voltage drop
across the junction, $\eta=0$, the energy of the unoccupied electron level will be
centered around the electrochemical potential of the right lead and moved away from
this value through the gate voltage. This ingredient will be crucial for the effect
addressed in this paper because the additional presence of the polaron shift will
then fix the energy of the occupied (charged) state below the electrochemical
potential of the right lead resulting in a blocked charged state under appropriate
parameter conditions. In the case of symmetric voltage drop, $\eta=0.5$, the energy
of the unoccupied electron level will be centered around the zero of the energy
resulting in a different scenario for what concerns the memory effect. In this paper
we consider the case of an asymmetric junction for which we show that the memory
effect occurs for small bias voltage in a wide range of the parameters entering the
model Hamiltonian (\ref{hamiltonian}). The symmetric situation can also give rise to
a hysteretic behaviour but only at finite bias voltages, making this case less
interesting for the memory effect addressed here.

\subsection{Spectral function, average charge and current}

To obtain physical information from the Hamiltonian (\ref{hamiltonian}) we use the
Green function method within the equation of motion (EOM) formalism.
This method is an alternative to the Green function techniques earlier applied to the
considered problem in
Refs.\,\cite{Galperin05nanolett,Ryndyk08preprint,Hewson74jjap,Hewson79jphysc}.
We start with the
retarded Green function for two generic operators $a$ and $b$ which is defined as
\begin{equation}\label{green_general}
G^{r}(t_1,t_2)=-i\Theta\left( t_1-t_2\right) \left\langle \left\lbrace a(t_1),b(t_2)\right\rbrace
\right\rangle \, ,
\end{equation}
where $ \left\langle,\right\rangle$ denotes a thermal average and $\{a,b\}=ab+ba$ is
the anti-commutator.
For the stationary case, the general expression (\ref{green_general}) reduces to an object with only one time argument,
\begin{equation}\label{green}
G^{r}(t)=-i\Theta\left( t\right) \left\langle \left\lbrace a(t),b\right\rbrace
\right\rangle \, .
\end{equation}
For a single-level system the retarded Green function in Eq.~(\ref{green}) is obtained by replacing the generic operators $a$ and $b$ by the electronic operators. For the Hamiltonian (\ref{hamiltonian}), this reads
\begin{equation}\label{green_vibron}
G^{r}(t)=-i\Theta\left( t\right) \left\langle \left\lbrace d(t),d^{\dagger}\right\rbrace
\right\rangle \, .
\end{equation}
From Eq.~(\ref{green_vibron}) one obtains the spectral function $A(\epsilon)$ of the system  through the expression
\begin{equation}\label{spectral function}
A(\epsilon)=-2 {\rm Im}G^r(\epsilon),
\end{equation}
where $G^{r}(\epsilon)$ is the Fourier transform of $G^{r}(t)$. The spectral function is the basic ingredient for obtaining the transport properties of the system such as average current and charge on the molecule. The expression for the current through the molecule is given by
\begin{equation}\label{current}
I=\frac{e\Gamma_L\Gamma_R}{\Gamma_L+\Gamma_R}
  \int^{+\infty}_{-\infty}A(\epsilon)[f^0_{L}(\epsilon-e\varphi_L)-f^0_{R}(\epsilon-e\varphi_R)]\frac{d\epsilon}{2\pi}
  \, ,
\end{equation}
where $f^0_{i}$ is the equilibrium Fermi function in the $i$-th lead. The tunneling
couplings to the right ($\Gamma_R$) and left ($\Gamma_L$) leads are
\begin{equation}
\Gamma_{i}(\epsilon)=2\pi\sum_{k}|V_{i,k}|^2\delta(\epsilon-\epsilon_{i,k}) \, ,
\end{equation}
where the matrix elements $V_{i,k}$ are assumed to be energy-independent (wide-band
limit). The full level broadening is given by the sum
\mbox{$\Gamma=\Gamma_L+\Gamma_R$}. Below $\Gamma_R$ and $\Gamma_L$ are assumed to be the
same.

The average charge (number of electrons), $n=\left\langle d^{\dagger}d\right\rangle$,
is given by
\begin{equation}\label{mean number}
n=\int^{+\infty}_{-\infty}A(\epsilon)f(\epsilon)\frac{d\epsilon}{2\pi} \, ,
\end{equation}
where $f(\epsilon)$ is the distribution function of electrons inside the molecule. For the
approximations used in this paper, we employ the same distribution function as in the
non-interacting case,
\begin{equation}\label{f}
  f(\epsilon)=\frac{\Gamma_Lf_L^0(\epsilon-e\varphi_L)+\Gamma_Rf_R^0(\epsilon-e\varphi_R)}
  {\Gamma_L+\Gamma_R} \, ,
\end{equation}
because we are focusing on the case of intermediate molecule-lead coupling. Fast
tunneling into and out of to the molecule makes plausible the assumption that the
electrons are in strong a non-equilibrium situation and can then be decribed via
Eq.~(\ref{f}) that is obtained assuming only elastic processes. Moreover, within the
first approximation that we discuss in the paper,  it is possible to obtain the
distribution function (\ref{f}) analytically by calculating the lesser Green
function of the problem or applying the Hartree approximation directly to the
Hamiltonian as in Ref.\,\cite{Galperin05nanolett}.

For a more complete explanation of the basic formulas introduced here the reader is
referred to Refs.~\cite{Haug96book,Bruus04book,Ryndyk08review}.

\section{Equation-of-motion method}

\subsection{General formalism}

The equations of motion for non-equilibrium Green functions are obtained from the
Heisenberg equation for a Heisenberg operator $a(t)$,
\begin{equation}\label{HeisenbergEq}
i\frac{\partial a}{\partial t}=\left[a,\hat H\right]_-=a\hat H-\hat H a \, .
\end{equation}
Here and below all Hamiltonians are assumed to be time-independent because we
consider the  stationary problem, the applied voltages enter the problem only as
time-independent parameters, changing the position of the molecular energy level and
the electrochemical potentials in the left and the right lead. With
Eq.~(\ref{HeisenbergEq}) the time derivative of the Green function,
Eq.~(\ref{green}), reads
\begin{equation}\label{green_derivative}
i\frac{\partial G^{r}(t)}{\partial t}=\delta(t)\left\langle \left\lbrace a(t),b\right\rbrace \right\rangle-i\Theta\left(
 t\right) \left\langle \left\lbrace \left[ a,H\right] (t),b\right\rbrace \right\rangle \, .
\end{equation}
After performing a Fourier transform of  Eq.~(\ref{green_derivative}) we obtain
\begin{equation}\label{green_fourier}
(\epsilon+i\eta)\left\langle \left\langle a,b\right\rangle
\right\rangle=\left\langle \left\lbrace a,b\right\rbrace
 \right\rangle +\left\langle \left\langle \left[ a,H\right] ,b\right\rangle \right\rangle \, ,
\end{equation}
where $\left\langle \left\langle,\right\rangle \right\rangle$ indicates the Fourier
transform of a given Green function. Equation (\ref{green_fourier}) is the starting
point for the EOM method. By applying successively a time derivative there are new
high-order Green functions appearing. The idea of the EOM method is thereby to
truncate this iterative process at some point making the mean-field like
approximation of the highest-order Green function through lower-order functions, in
order to obtain a closed set of equations. In our case we start from the Hamiltonian
(\ref{hamiltonian}) and consider the first-oder equation for the function
$\left\langle \left\langle d,d^{\dagger}\right\rangle \right\rangle$ and
second-order equations for the functions $\left\langle \left\langle
xd,d^{\dagger}\right\rangle \right\rangle$ and $\left\langle \left\langle
pd,d^{\dagger}\right\rangle \right\rangle$.

\subsection{EOM method for the single-level electron-vibron Hamiltonian}

The EOM method for Hamiltonian (\ref{hamiltonian}) generates the expressions
\begin{eqnarray} \label{equations 1}
(\epsilon+i\eta)\left\langle \left\langle d,d^{\dagger}\right\rangle \right\rangle
& = & 1+\tilde{\epsilon}_{0}\left\langle \left\langle d,d^{\dagger}\right\rangle
\right\rangle+\lambda\left\langle \left\langle xd,d^{\dagger}\right\rangle
\right\rangle
\nonumber \\
& & \qquad + \sum_{i,k}V^{*}_{i,k}\left\langle\left\langle
c_{i,k},d^{\dagger}\right\rangle \right\rangle \, , \\
(\epsilon+i\eta-\epsilon_{i,k})\left\langle \left\langle
c_{i,k},d^{\dagger}\right\rangle \right\rangle
& = &
V_{i,k}\left\langle \left\langle
d,d^{\dagger}\right\rangle \right\rangle \, .
\label{set zero}
\end{eqnarray}
The equation for $\left\langle\left\langle c_{i,k},d^{\dagger}\right\rangle
\right\rangle$ is closed (including only the function $\left\langle \left\langle
d,d^{\dagger}\right\rangle \right\rangle$). By substituting Eq.~(\ref{set zero})
into Eq.~(\ref{equations 1}) and introducing the self-energy $\Sigma$ of the leads
through
\begin{equation}\label{self-energy}
\Sigma=\sum_{i,k}\frac{|V_{i,k}|^2}{\epsilon+i\eta-\epsilon_{i,k}}\, ,
\end{equation}
we obtain eventually
\begin{equation}\label{EqG}
  \left(\epsilon+i\eta-\tilde{\epsilon}_{0}-\Sigma\right)\left\langle \left\langle d,d^{\dagger}\right\rangle
  \right\rangle=1+\lambda\left\langle \left\langle xd,d^{\dagger}\right\rangle\right\rangle \, .
\end{equation}
The last term, describing the interaction between electron and vibron, has to be
truncated at this level or found from  higher-order equations and then truncated
at  a higher level of approximation. The lead self-energy will be used below in the wide-band
approximation $\Sigma(\epsilon)=-i\Gamma$.

\subsection{Self-consistent Hartree approximation}

The simplest way to close Eq.~(\ref{EqG}) is to perform the truncation by
approximating
\begin{equation}\label{approx 1}
\left\langle \left\langle xd,d^{\dagger}\right\rangle \right\rangle \approx
\left\langle x\right\rangle \left\langle \left\langle d,d^{\dagger}\right\rangle \right\rangle \, .
\end{equation}
Then we obtain immediately for the Green function
\begin{equation}
\label{Gr}
G^r_{H}(\epsilon)=\left\langle \left\langle d,d^{\dagger}\right\rangle \right\rangle =
\frac{1}{\epsilon-\tilde{\epsilon}_{0}- \lambda \left\langle x\right\rangle+i\Gamma}\, .
\end{equation}
Here the quantity $\left\langle x\right\rangle$ remains to be
calculated. To this end we compute, respectively,
the time derivatives of the $x$-operator,
\begin{equation}\label{x derivative}
i\frac{\partial x}{\partial t} =\left[ x,H\right] =\omega_{0}p \, ,
\end{equation}
and the $p$-operator,
\begin{equation}\label{p derivative}
i\frac{\partial p}{\partial t} = \omega_{0} x+2\lambda d^{\dagger}d \, .
\end{equation}
Upon combining Eqs.~(\ref{x derivative}) and (\ref{p derivative}) we get
\begin{equation}\label{x 2derivative}
-\frac{\partial^2x}{\partial t^2}=\omega_{0}^2 x+2\lambda \omega_{0} d^{\dagger}d\, .
\end{equation}
In the stationary case addressed, Eq.~(\ref{x 2derivative}) yields
a direct connection between the "position" of the vibron and the
particle number in the dot:
\begin{equation}\label{position}
\left\langle x\right\rangle =-2\frac{\lambda}{\omega_{0}} \left\langle
d^{\dagger}d\right\rangle =-2\frac{\lambda}{\omega_{0}} n \, .
\end{equation}
In view of Eq.~(\ref{Gr}), we finally obtain for the spectral function (\ref{spectral function}) the following self-consistent expression
,
\begin{equation}\label{green 1}
A(\epsilon)=\frac{2 \Gamma}{\left(
\epsilon-\tilde{\epsilon}_{0}+2\frac{\lambda^2}{\omega_{0}}n\right) ^2+\Gamma^2}\, .
\end{equation}
This result is equivalent to the one  obtained earlier in
Refs.~\cite{Hewson74jjap,Hewson79jphysc,Galperin05nanolett} using alternative
approaches. The same spectral function can be found if one takes the self-energy in
Hartree approximation.

\subsection{Second approximation}

In the first approximation above, the self-consistent Hartree treatment,
fluctuations of the particle number $n$ and the vibron coordinate $x$
are completely neglected. In order to go one step further and estimate
possible corrections, we start from the generated equations for the
second-order Green functions,
$\left\langle \left\langle xd,d^{\dagger}\right\rangle \right\rangle$
and
$\left\langle \left\langle pd,d^{\dagger}\right\rangle \right\rangle$,
\begin{eqnarray} \label{xdd}
(\epsilon+i\eta-\tilde{\epsilon}_{0})\left\langle \left\langle xd,d^{\dagger}\right\rangle \right\rangle
\nonumber \\ =\left\langle x\right\rangle+ \omega_0\left\langle \left\langle pd,d^{\dagger}\right\rangle
\right\rangle+\lambda\left\langle \left\langle x^2d,d^{\dagger}\right\rangle
\right\rangle +\sum_{i,k}V^{*}_{i,k}\left\langle\left\langle
xc_{i,k},d^{\dagger}\right\rangle \right\rangle \, , \\ \label{pdd}
(\epsilon+i\eta-\tilde{\epsilon}_{0})\left\langle \left\langle pd,d^{\dagger}\right\rangle \right\rangle
\nonumber \\ =\omega_0\left\langle \left\langle xd,d^{\dagger}\right\rangle
\right\rangle+\lambda\left\langle \left\langle pxd,d^{\dagger}\right\rangle
\right\rangle +\sum_{i,k}V^{*}_{i,k}\left\langle\left\langle
pc_{i,k},d^{\dagger}\right\rangle \right\rangle \, .
\end{eqnarray}
The second approximation that we consider here is based on the factorization
\begin{eqnarray}\label{approx 2}
\left\langle \left\langle x^2d,d^{\dagger}\right\rangle \right\rangle
\approx \left\langle x\right\rangle \left\langle \left\langle xd,d^{\dagger}\right\rangle
\right\rangle, \nonumber \\
\left\langle \left\langle pxd,d^{\dagger}\right\rangle \right\rangle
\approx \left\langle x\right\rangle \left\langle \left\langle pd,d^{\dagger}\right\rangle
\right\rangle +2\left\langle \left\langle d,d^{\dagger}\right\rangle \right\rangle, \nonumber \\
\left\langle \left\langle xc_{i,k},d^{\dagger}\right\rangle \right\rangle
\approx \left\langle x\right\rangle \left\langle \left\langle c_{i,k},d^{\dagger}\right\rangle \right\rangle, \nonumber \\
\left\langle \left\langle pc_{i,k},d^{\dagger}\right\rangle \right\rangle\approx 0 \, .
\end{eqnarray}
The corresponding set of equations reads
\begin{eqnarray}
(\epsilon+i\eta-\tilde{\epsilon}_0-\Sigma)\left\langle \left\langle d,d^{\dagger}\right\rangle \right\rangle&=
&1+\lambda\left\langle \left\langle x d,d^{\dagger}\right\rangle \right\rangle,
\nonumber\\
(\epsilon+i\eta-\tilde{\epsilon}_0-\lambda \left\langle x \right\rangle)\left\langle \left\langle x d,d^{\dagger}\right\rangle \right\rangle&=
&\left\langle x\right\rangle +\omega_{0}\left\langle \left\langle pd,d^{\dagger}\right\rangle \right\rangle+
\left\langle x \right\rangle \Sigma\left\langle \left\langle d,d^{\dagger}\right\rangle \right\rangle,\nonumber\\
(\epsilon+i\eta-\tilde{\epsilon}_0-\lambda \left\langle x \right\rangle)\left\langle \left\langle p d,d^{\dagger}\right\rangle \right\rangle&=
&\omega_{0}\left\langle \left\langle xd,d^{\dagger}\right\rangle \right\rangle+
2\lambda \left\langle \left\langle d,d^{\dagger}\right\rangle \right\rangle \label{terzo set},
\end{eqnarray}
from which
we obtain the second approximation for the Green function,
\begin{equation}\label{green2} \left[G^r(\epsilon)\right]^{-1}=
 \left[G^r_{H}(\epsilon)\right]^{-1}-\left( \lambda\left\langle x\right\rangle +\frac{\lambda^2}{\omega_0}\right)
 \frac{\Omega}{\Delta-\Omega}
 \, ,
\end{equation}
where we introduced $\Delta=\epsilon+i\eta-\tilde{\epsilon}_0$,
$\Omega=\frac{\omega_{0}^{2}}{\Delta-\lambda\left\langle x\right\rangle}$, and
$G^r_{H}$ is given by the Green function obtained in the Hartree approximation,
Eq.(\ref{Gr}). After inserting the expression (\ref{position}) for the level
population $\left\langle x\right\rangle$,  Eq.~(\ref{green2}) reads
\begin{equation}\label{green2_new}
\left[G^r(\epsilon)\right]^{-1}=
 \left[G^r_{H}(\epsilon)\right]^{-1}-\left(1-2n\right) \frac{\lambda^2}{\omega_0}\frac{\Omega}{\Delta-\Omega}
 \, .
\end{equation}
We then calculate the spectral function (\ref{spectral function}) and the average
number of electrons, (\ref{mean number}).
The self-consistent
calculation is performed following the chain $G^r(\epsilon)\rightarrow
A(\epsilon)\rightarrow n\rightarrow G^r(\epsilon)$.

\begin{figure}
\begin{center}
\epsfxsize=0.49\hsize \epsfbox{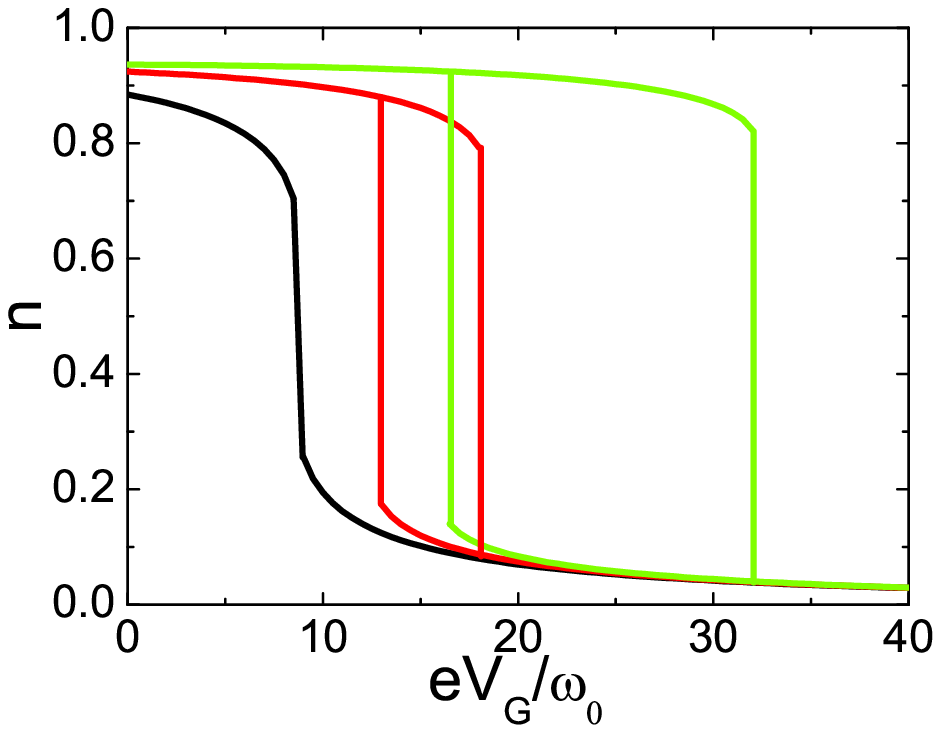} \epsfxsize=0.49\hsize
\epsfbox{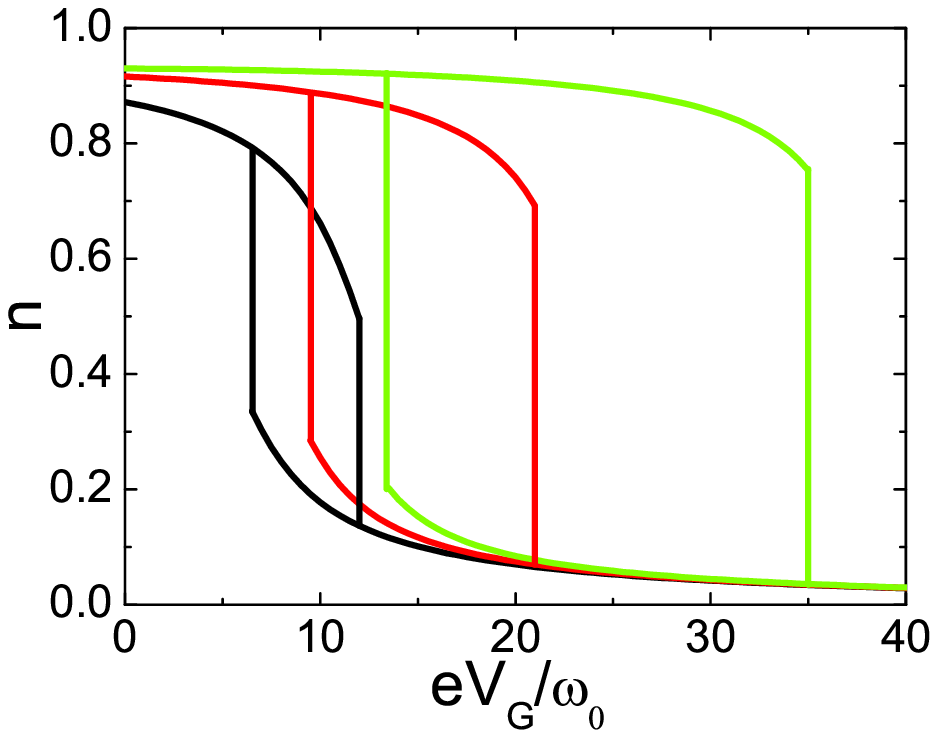} \caption{(Color) Bistable behavior of the
level population $n$ in the
self-consistent Hartree  approximation (left) and second approximation
(right) as a function of {\em gate} voltage for different electron-vibron
interaction strength $\lambda=3\omega_0$ (black),
$\lambda=4\omega_0$ (red), $\lambda=5\omega_0$ (green), the other parameters are
$\Gamma=5\omega_0$, $T=0.25\omega_0$, $\eta = 0$ and $V_B=0$.}
\label{GraphFirstGate.eps}
\end{center}
\end{figure}

Before entering into the discussion of the calculated quantities, we consider the
structure of the Green functions obtained in the two different approximations:
\begin{itemize}

\item In the limit where $\Omega\rightarrow 0$ the second approximation
    reduces to the first one. Although the EOM method is
    not a systematic expansion, it tells us that the second approximation
    consistently extends the first one and reproduces it in a limiting case.

\item The second term on the right site of Eq.~(\ref{green2_new}) represents an additional shift with respect to Eq.~(\ref{Gr}). In the case of very small frequencies $\omega_0$, Eq.~(\ref{green2_new}) reduces to $\left[G^r_{H}(\epsilon)\right]^{-1}-\left(1-2n\right) \lambda^2\frac{\omega_0}{\Delta(\Delta-\lambda\left\langle x\right\rangle) }$, involving the first term of a series expansion in $\omega_0$.
\end{itemize}

\section{Results, discussion and conclusions}

Starting from the expressions derived above within the first (self-consistent Hartree)
and second approximation we have performed numerical simulations for the
average population and current in the molecular junction as a function of
gate and bias voltage for the different parameters $\lambda, \Gamma$ and $T$
entering the model Hamiltonian. Furthermore we compare the two approximations.
The underlying nonlinear equations give rise to bistability in the level
population thereby enabling memory effects and affecting the current.
Below we analyze in detail the parameter ranges and conditions for
memory effects to occur.
We focus here on the case of an asymmetric junction, $\eta = 0$, while
keeping the coupling to the leads symmetric, $\Gamma_L=\Gamma_R=\Gamma$,
to reduce the parameter space.

First we investigate the gate-voltage dependence of the level population.
In Fig.~\ref{GraphFirstGate.eps}
it is clearly seen that bistability takes place only at larger values of the
electron-vibron coupling $\lambda$. The critical values at which bistability occurs
and disappears depend on the coupling  $\Gamma$ to the leads and on temperature $T$;
we discuss this parameter dependence below. Note that at large values of $\lambda$
the level population of the stable states is close to 0 and 1; thus these two memory
states are well distinguishable in charge.

\begin{figure}
\begin{center}
\epsfxsize=0.49\hsize \epsfbox{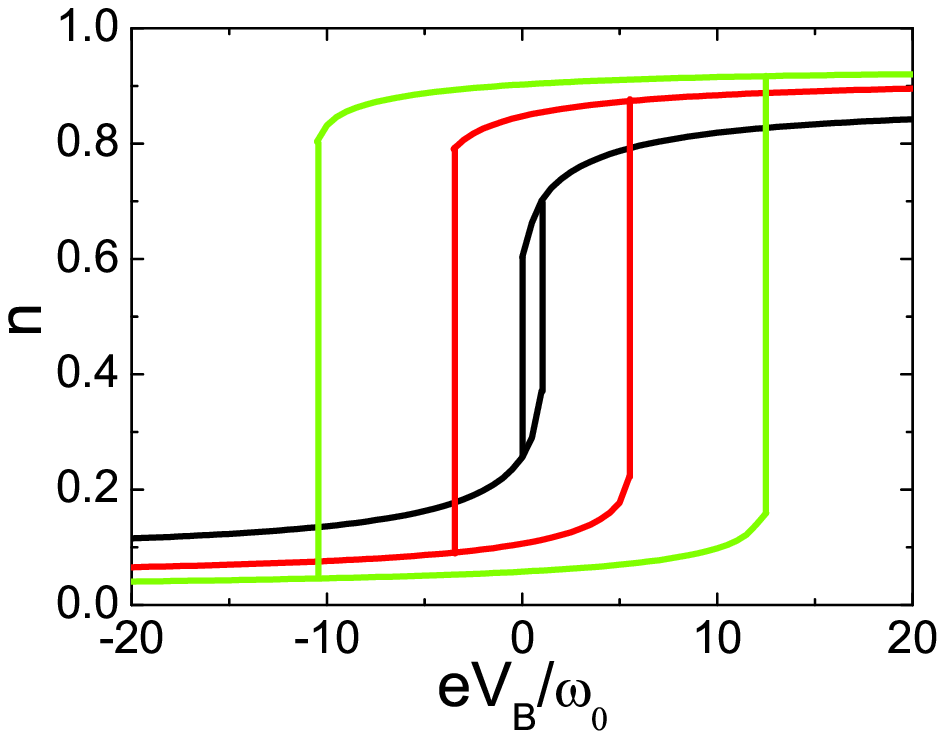}
\epsfxsize=0.49\hsize \epsfbox{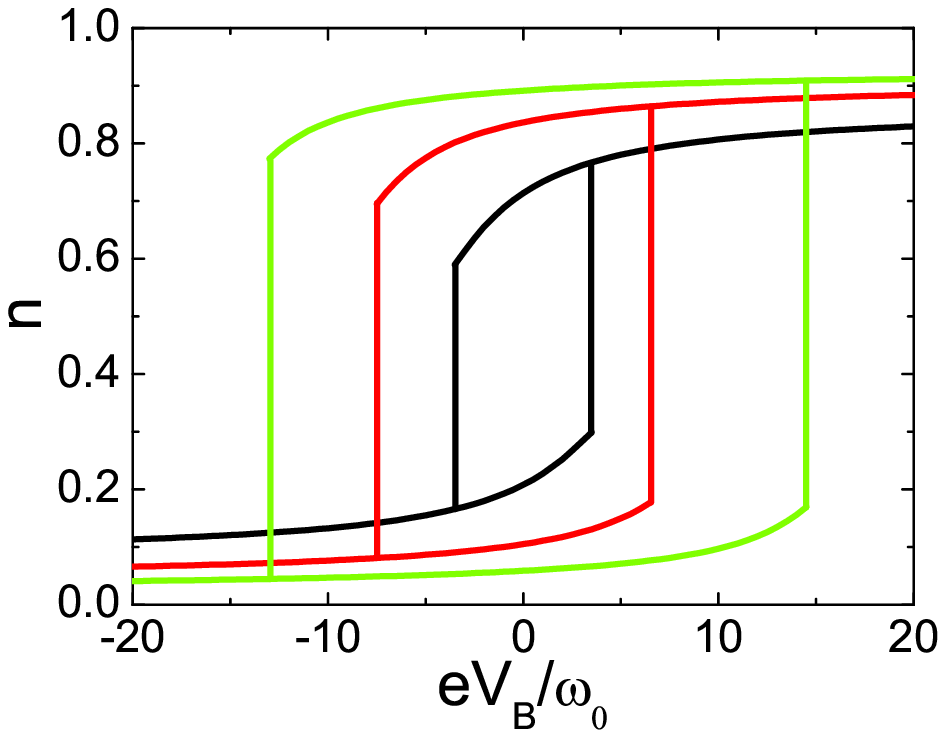}
\caption{(Color) Level population $n$ in the self-consistent Hartree
 approximation (left) and the second approximation (right) as a function of
 normalized {\em bias} voltage for different electron-vibron
 interaction strength $\lambda=3\omega_0$ (black), $\lambda=4\omega_0$ (red),
$\lambda=5\omega_0$ (green), the other parameters are $\Gamma=5\omega_0$,
$T=0.25\omega_0$, $\eta=0$ and $V_G=\frac{\lambda^2}{\omega_0}$.}
\label{GraphFirstBias.eps}
\end{center}
\end{figure}

A comparison between the left and right panel of Fig.~\ref{GraphFirstGate.eps},
representing the two different levels of approximations in the EOM method, shows
that the self-consistent Hartree treatment  underestimates the parameter range where
bistability occurs: the critical value of $\lambda$ for the occurence of bistable
behavior is close to $3 \omega_0$ (see left panel). At this value a bistable regime
has already developed for the second approximation. This can be partially understood
taking into account the additional term appearing in the Green function for the
second approximation, Eq.~(\ref{green2_new}). This term increases the polaron shift
thereby enhancing the bistable behaviour in the second approximation.
Fig.~\ref{GraphFirstBias.eps} shows the bias-voltage dependence of the level
population. In this case a qualitative difference arises between electrostatically
symmetric ($\eta=0.5$) and asymmetric ($\eta=0$) junctions. For asymmetric junctions
both states are stable at zero voltage, and both charge states are easily
accessible. As we showed recently \cite{Ryndyk08preprint}, asymmetric junctions are
thus favourable, since they exhibit memory effects and hysteretic behavior at zero
bias, enabling controlled switching upon ramping the bias voltage. For symmetric
junctions hysteresis is expected only at finite bias voltage (nonequilibrium
bistability \cite{Galperin05nanolett}), and hence only a single stable state exists
at zero bias. Furthermore, at finite voltage the level is only partially occupied by
tunneling electrons. These two features render the symmetric system less suited for
a memory setup.

\begin{figure}[t]
\begin{center}
\epsfxsize=0.49\hsize \epsfbox{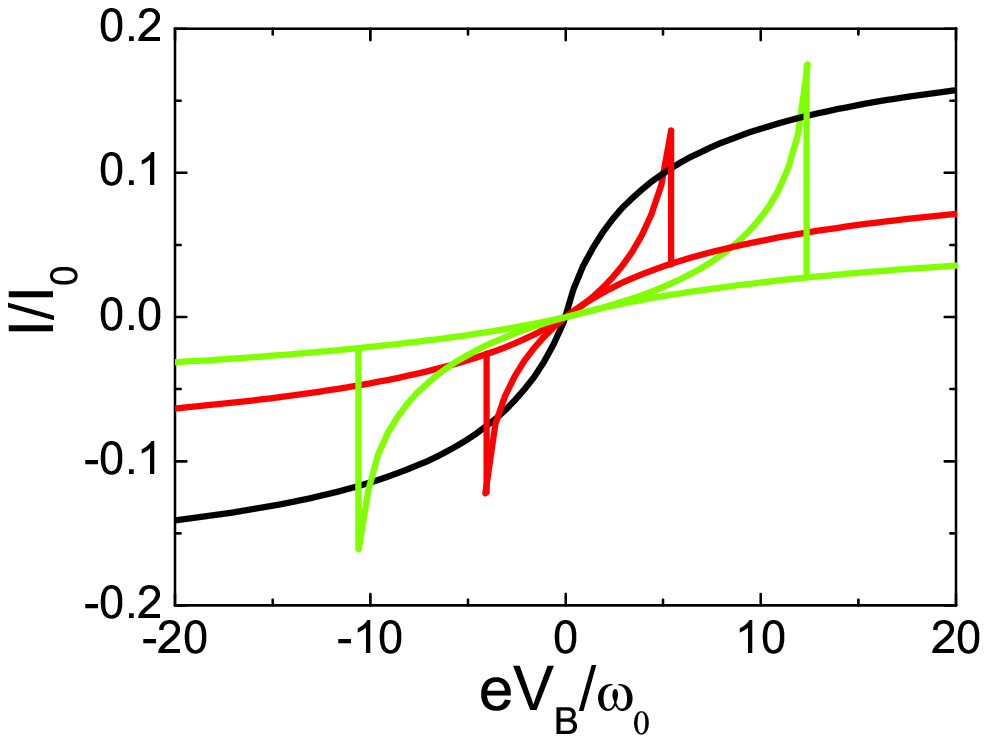}
\epsfxsize=0.49\hsize \epsfbox{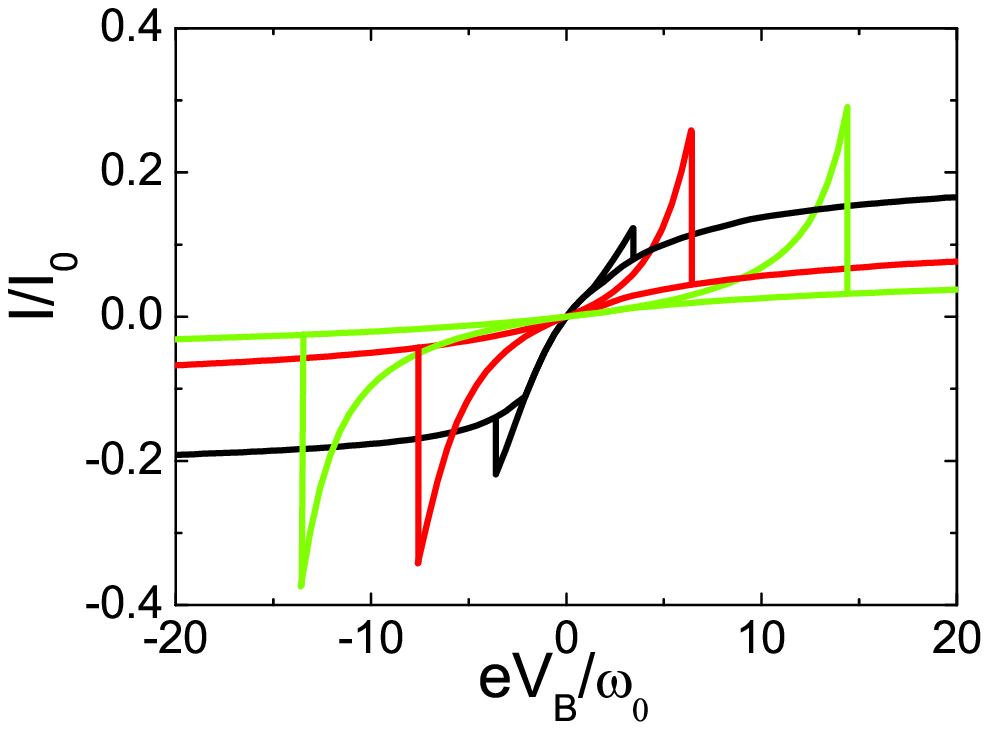}
\caption{(Color) Current normalized to $I_0 =\Gamma e/4$ versus bias voltage for the
self-consistent Hartree
 approximation (left) and the second approximation (right) for different electron-vibron
interaction strength $\lambda=3\omega_0$ (black), $\lambda=4\omega_0$ (red),
$\lambda=5\omega_0$ (green), for the same parameters as in Fig.~\ref{GraphFirstBias.eps}.}
\label{GraphFirstCurrent.eps}
\end{center}
\end{figure}

We further display  in Fig.~\ref{GraphFirstCurrent.eps}
the current-voltage characteristics, which reflects the switching
behavior of the system. The characteristic feature
is a current jump at the bias voltage value where recharging sets in. This behavior
can be used to test the state of the system and as readout.

\begin{figure}[b]
\begin{center}
\epsfxsize=0.49\hsize \epsfbox{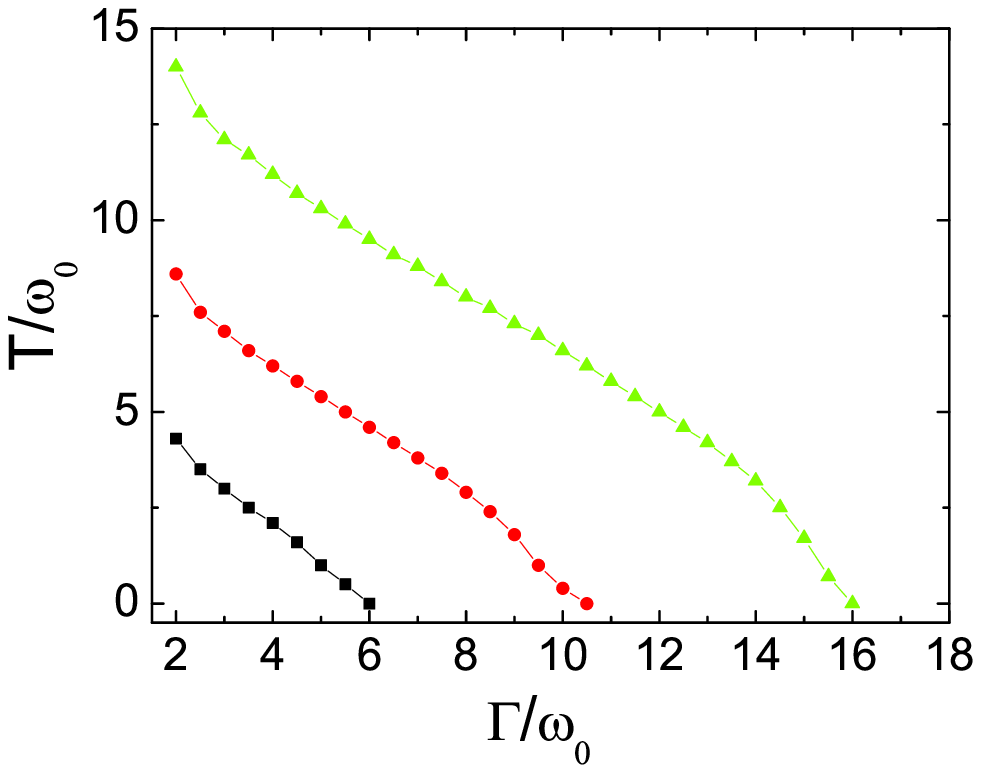}
\epsfxsize=0.47\hsize \epsfbox{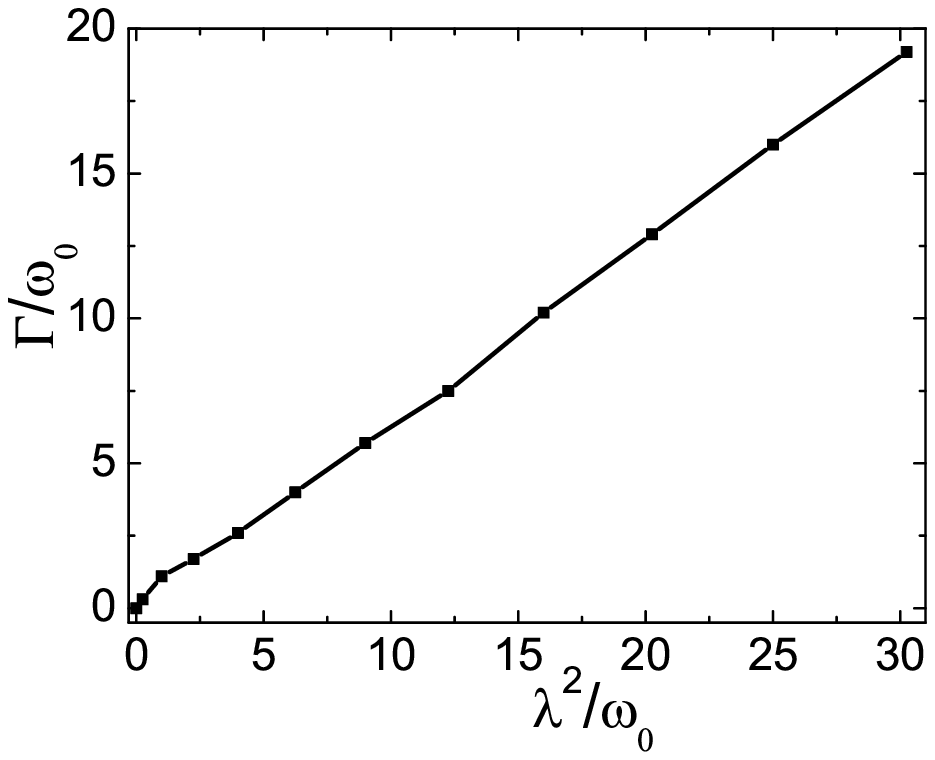}
\caption{(Color) "Phase diagram"  depicting the boundaries between parameter regimes
of bistable memory (below the threshold lines) and single-valued states (above threshold).
Left panel: $\Gamma-T$ parameter plane
at different electron-vibron interaction strength $\lambda=3\omega_0$ (black),
$\lambda=4\omega_0$ (red), $\lambda=5\omega_0$ (green); right panel:
 $\lambda-\Gamma$ paramter plane for $T=0.25\omega_0$.}
\label{GraphPhaseG-T.eps}
\end{center}
\end{figure}

Finally, we depict in Fig.~\ref{GraphPhaseG-T.eps} "phase diagrams" showing the
boundaries between the parameter regions where bistable memory states (below the
boundaries in the two panels of Fig.~\ref{GraphPhaseG-T.eps}) and single-valued
states exist. The left panel of  Fig.~\ref{GraphPhaseG-T.eps} shows the  $\Gamma-T$
parameter plane. The curve separating single-valued and bistable states can be
roughly approximated by the condition $\Gamma+T = c(\lambda)$, where $c(\lambda)$ is
extracted to be $c(\lambda)\approx \lambda^{1.7}$. This means that either thermal or
quantum tunneling broadening suppresses the hysteresis. The disappearance of the
memory effect at high temperature is due to enhanced electron tunneling into the
higher-energy state. We note that in the analysis of the temperature dependence the
effect of vibronic coordinate fluctuations is neglected that can be relevant close
to the threshold between single-valued and bistable  regime. The right panel in
Fig.~\ref{GraphPhaseG-T.eps}, displaying the $\lambda^{2}-\Gamma$ parameter plane,
shows on the one hand that bistable behaviour requires, for growing $\Gamma$,
increasing electron-vibron coupling $\lambda$. The boundary between the two regimes
is approximately a straight line in the $\lambda^2-\Gamma$ plane. Hence the
condition for finding the memory effect is given by $\Gamma \leq 0.63 \lambda^{2}/
\omega_0$. This clearly shows that for the appearence of the memory effect at low
temperature, the two energy scales to be compared are the level broadening $\Gamma$
and the polaron shift $\lambda^{2}/\omega_{0}$. A further important conclusion is
that the memory effect is suppressed for large coupling to the leads. Since, on the
other hand, larger coupling favours fast information writing and reading and also
can additionally suppress effects from  quantum tunneling between states, the
problem arises to find optimal parameters for utilizing the memory effect. This will
be the subject of future work.

To conclude, we considered a {\em charge-memory effect} and switching phenomena
within a single-level polaron model of a molecular junction in the framework of the
equation-of-motion approach to the nonequilibrium Green function theory at different
levels of approximation. Electrostatically symmetric and asymmetric junctions show
qualitatively different bistability behavior. In the latter case, controlled
switching of the molecule is achieved by applying finite voltage pulses. We showed
that bistability takes place for sufficiently large elelctron-vibron coupling for a
wide range of further parameters such as molecule-lead coupling and temperature and
performed a systematic analysis of this parameter dependence by computing phase
diagrams for the memory effect.  We focussed on the investigation of bistability for
stationary states, leaving the problem of time-dependent fluctuations and
fluctuation-induced switching as an open question to be investigated in future
research.

\section{Acknowledgments}
We acknowledge fruitful discussions with J. Repp and thank one of the referees for
useful hints. This work was funded by the Deutsche Forschungsgemeinschaft within
Priority Program SPP 1243, Collaborative Research Center SFB 689 (D.A.R.) and within
the international collaboration "Single molecule based memories" CU~44/3-2 (D.A.R.).

\end{document}